\documentclass[%
 reprint,
 amsmath,amssymb,
 aps,
]{revtex4-2}

\usepackage{graphicx}
\usepackage{dcolumn}
\usepackage{bm}

\begin{document}

\preprint{APS/123-QED}

\title{Energy landscape transformation of Ising problem with invariant eigenvalues for quantum annealing}

\author{T. Fujii}
 \email{Toru.Fujii@nikon.com}
\author{K. Komuro}
\author{Y. Okudaira}
\affiliation{%
 Nikon Corporation, 2-15-3, Konan, Minato-ku, 108-6290 Tokyo, Japan 
}%

\author{R. Narita}
\author{M. Sawada}
\affiliation{
 Nikon Systems Inc., 1-6-3, Nishioi, Shinagawa-ku, 140-0015 Tokyo, Japan
}%

\date{\today}

\begin{abstract}
Quantum annealing tends to be more difficult as the energy landscape of the problem becomes complicated with many local minima. We have found a transformation for changing the energy landscape that swaps the eigenvalues and paired states without changing the eigenvalues of the instance at all. The transformation is basically a partial recombination of the two-spin interaction coefficient ``${J_{ij}}$" and the longitudinal magnetic field interaction coefficient ``${h_i}$". The Hamming distance corresponding to a barrier between the states changes by the transformation, which in turn affects the ground state convergence. In the quantum annealing simulation results of a small number of spin instances, the annealing time was shortened by several orders of magnitude by applying the transformation. In addition, we also obtained a result using a D-Wave quantum annealer, which also showed a big improvement in the ground state convergence.
\end{abstract}

\maketitle


Quantum annealing (QA) \cite{Kadowaki98,Albach18} is a promising method for solving computationally difficult problems using adiabatic changes, and many examples of solutions using actual machines have already been reported. It has also been shown theoretically that the time required for convergence is shorter than simulated annealing  \cite{Morita07}. Quantum logic gate devices are in the “noisy intermediate-scale quantum” (NISQ) era and quantum devices of quantum annealer are no exception. It is not realistic to perform QA for a time that is theoretically sufficient for convergence. Instead, various speed-up methods have been proposed to overcome noise limits in both QA and NISQ gate-based devices  \cite{Crosson21}.

It is also well known that there is a difficulty level in the Ising problem. Nonstoquastic Hamiltonians  \cite{Nishimori17} that positively use the quantum property of QA have also been shown to be effective for difficult problems \cite{Crosson14}. It has been shown that the Ising problem with randomly generated coefficients becomes NP-hard \cite{Barahona82}. However, the probability that it is a difficult problem for nonstoquastic QA was low, when the Ising coefficients are generated randomly as shown by Crosson et al. When the system to be handled becomes large, the Ising spin glasses of the 2D vertical magnetic field model and the 3D model also become NP-hard. In contrast, if the spin number is small and the Ising coefficients are randomly generated, it becomes an easy problem \cite{Crosson14}.

The difficulty of the Ising problem is in its complex energy landscape with many local minima  \cite{MM09}.  A small Hamming distance of excited states to the closest ground state tends to create an easy Ising problem, and a long Hamming distance tends to be the opposite \cite{Boixo14}. Analysis of related barrier tunneling problems have also been performed \cite{Brady16}. As the original description of the problem difficulty, the case of geometrical frustration caused by antiferromagnetism is often used  \cite{Moessner06}. Following this model, and by deliberately creating conundrums with small number of spins, we discuss the effects of our method, the nature of the conundrums, and speculate about the phenomena in large-scale problems.

A QA problem of an $n$ spin system can be expressed as the following equation (1) to (3) \cite{Kadowaki98}:

\begin{eqnarray}
H(s)=(1-s)H_B+sH_P
\label{eq:one}.
\end{eqnarray}
${H_B}$ is a transverse magnetic field added by the Pauli $x$-matrix $\sigma_i^x$, as shown in Eq. (2),

\begin{eqnarray}
H_B=\sum_{i=0}^{n-1}\sigma_i^x\ 
\label{eq:two}.
\end{eqnarray}
Here, the problem Hamiltonian is expressed by the Ising coefficients as

\begin{eqnarray}
H_P = \sum_{i<j} J_{ij}\sigma_i^z\sigma_j^z +  \sum_i h_i\sigma_i^z
\label{eq:three}.
\end{eqnarray}
Equation (3) consists of the two-spin interaction coefficient ${J_{ij}}$ and the longitudinal magnetic field interaction coefficient ${h_i}$. Problems such as traveling salesman problem (TSP) require a regularization term as a penalty term, but the Ising coefficients are uniquely calculated as in ${H_P}$ in Eq. (1) except for the degree of freedom of the regularization term. 

We show a method that can change the difficulty of the Ising problem itself, with ramifications to the total required annealing time for ground state convergence. An important parameter for varying difficulty is the Hamming distance between the energy gap and the low energy state. In order to clearly show the effect, the difficult Ising problem we evaluate is generated  by adjusting the 4 spin Ising coefficients according to the policy of Boixo et al. \cite{Boixo14}, so that the problem includes the case in which there are multiple excited states with small gaps from the ground state and the Hamming distance is long.

Figure 1(a) and (b) show an example in which the annealing time first depends on the energy gap. The time evolution of the quantum state in QA is derived by numerically integrating the discretized Schr\"odinger equation with standard annealing schedules. The vertical axis is the convergent probability of final state, and the horizontal axis is annealing total time. All Ising coefficients for QA in this Letter are shown in Table 1.

\begin{figure}[!b]
  \centering
  \includegraphics[width=1\columnwidth]{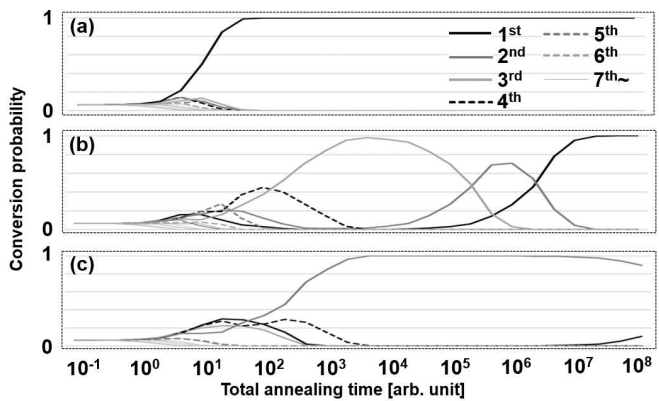}
  \caption{Simulation result of each state conversion versus total QA time. Ising coefficients shown above were determined not to degenerate each eigenvalue, (see Table 1). Each eigenvalue and order is shown in Table 2.
  (a) Ising problem of large energy gap of 1.4, (b) small energy gap of 0.005, whose state is located Hamming distance of 3 far from ground state, (c) small energy gap of 0.01, whose state is located with Hamming distance of 4, which is the longest distance of $4$ spin Ising problem. Vertical axis: Convergent probability of final state, Horizontal axis: Annealing time.}
  \label{fig:image2}
\end{figure}

In Fig. 1(a) the left end where the annealing time is very short, the influence of the transverse magnetic field does not disappear due to annealing, and each state probability is almost the same. As the annealing time is extended, the final state probability changes as shown on the right.

While the energy gap to the first excited state of the Ising problem in Fig. 1(a) is 1 or more, that in Fig. 1(b) is as small as 0.005, and the Hamming distance is as far as 3. Furthermore, in Fig. 1(b), the energy difference is close to 0.015 even in the state $[-+-+]$ where the Hamming distance is the farthest from the ground state $[+-+-]$. Therefore, the Boixo condition is satisfied. The annealing time of Fig. 1(b) is 200,000 times longer than that of Fig. 1(a). Interestingly, the ground state does not converge until first the probability of existence of multiple excited states rises and falls once. Especially when the Hamming distance is 4, the farthest from the ground state, the maximum state probability rises to near 1.

In Fig. 1(c), the energy gap to the first excited state is slightly widened to 0.01, and the Hamming distance is set to 4, the farthest from the ground state. In the total annealing time, which is almost converged in Fig. 1(c), the first excited state has a very high population, and the transition to the ground state starts from $T=10^7$ when the annealing of Fig. 1(b) was stopped. When there are many low-energy states, it takes time to transit between the states, and the Hamming distance \cite{Fahri02,Boixo14}, which acts like a tunnel barrier, is considered to have a large effect on the required total annealing time.

It is well known \cite{Sullivan08} that the original Hamiltonian Eq. (3) can be modified to

\begin{eqnarray}
\tilde{H} = \sum_{\substack{i<j\\i,j=0}}^{n} \tilde{J}_{ij}\sigma_i^z\sigma_j^z = \sum_{\substack{i<j\\i,j=0}}^{n-1} J_{ij}\sigma_i^z\sigma_j^z+\sum_{i=0}^{n-1}h_i\sigma_i^z\sigma_n^z
\label{eq:four}.
\end{eqnarray}
where $\sigma_n^z$ is an ancilla spin and $\tilde{J}_{in}=h_i$, $\tilde{J}_{ij}=J_{ij}$ $(i,j\in{[0,n-1]})$.  Since the parity transformed Hamiltonian $H^{\prime}=\sum_{i<j}J_{ij}\sigma_i^z\sigma_j^z-\sum_{i}h_i\sigma_i^z$ has the same eigenvalues as the original $H$,  all eigenstates of Eq (4) are doubly degenerated, which guarantees a state $|s_{0},\cdots ,s_{n} \rangle$ and its parity transformed state $|-s_{0},\cdots ,-s_{n} \rangle$ has the same eigen energy. 
Using this degeneracy, it is possible to fix one spin polarity of any $\sigma_k^z (k=0.\cdots ,n)$ without changing the eigenvalues.
Meanwhile spin swapping $\mathcal{S}_{nk}$ gives a Hamiltonian:

\begin{eqnarray}
\mathcal{S}_{nk}[\tilde{H}] &\equiv& \sum_{\substack{i,j \neq{k}\\i,j=0}}^{n-1} J_{ij}\sigma_i^z\sigma_j^z+\sum_{\substack{i\neq{k}\\i=0}}^{n-1} J_{ik}\sigma_i^z\sigma_n^z\nonumber \\
&&+\sum_{\substack{i\neq{k}\\i=0}}^{n-1} h_{i}\sigma_i^z\sigma_k^z+h_{k}\sigma_{n}^z\sigma_{k}^z
\label{eq:five}
\end{eqnarray}
whose eigenvalues are also the same as the original ones.
Fixing the ancilla spin in Eq. (5) to be $\sigma_n^{z}=1$, we have

\begin{eqnarray}
\mathcal{T}_{k}[\tilde{H}] &\equiv& \sum_{\substack{i,j \neq{k}\\i,j=0}}^{n-1} J_{ij}\sigma_i^z\sigma_j^z+\sum_{\substack{i\neq{k}\\i=0}}^{n-1} J_{ik}\sigma_i^z\nonumber \\
&&+\sum_{\substack{i\neq{k}\\i=0}}^{n-1} h_{i}\sigma_i^z\sigma_k^z+h_{k}\sigma_k^z
\label{eq:six}
\end{eqnarray}
in which the role of $h_i$ and $J_{ij}$ is replaced when $k \neq n$.
Since this transformation $\mathcal{T}$ may create $(n+1)$ types of energy landscapes including the original one, we call it the energy landscape transformation ``ELTIP''.

With an even anti-ferromagnetic Ising problem whose ${J_{ij}}$ are all positive, the non-zero ${h_{i}}$ can be replaced with ${J_{ij}}$ by the transformation. We can choose negative ${h_{i}}$ by the simultaneous spin-and-${h_i}$ inversion. Transformed negative $J_{ik}^{\prime}$ may remedy the difficulty of anti-ferromagnetic Ising problem.

\begin{table*}[tb]
\caption{\label{tab:table1}%
Ising coefficients of each figure. Bold characters are $p_0$-transformed coefficients.
}
\begin{ruledtabular}
\begin{tabular}{lcccccccccc}
\textrm{Fig.}&
\textrm{$J_{01}$}&
\textrm{$J_{02}$}&
\textrm{$J_{03}$}&
\textrm{$J_{12}$}&
\textrm{$J_{13}$}&
\textrm{$J_{23}$}&
\textrm{$h_{0}$}&
\textrm{$h_{1}$}&
\textrm{$h_{2}$}&
\textrm{$h_{3}$}\\
\colrule
1(a)&0.7&1.4&1.9&0.9&0.3&2.1&1.1&0.3&-1.5&-0.9\\
2(a)&\textbf{0.3}&\textbf{-1.5}&\textbf{-0.9}&0.9&0.3&2.1&1.1&\textbf{0.7}&\textbf{1.4}&\textbf{1.9}\\
1(b)&0.75&1.375&1.875&0.125&0.375&2.25&0.7375&0.49&-0.675&-0.42\\
2(b)&\textbf{0.49}&\textbf{-0.675}&\textbf{-0.42}&0.125&0.375&2.25&0.7375&\textbf{0.75}&\textbf{1.375}&\textbf{1.875}\\
1(c)&0.75&1.375&1.875&0.125&0.375&2.25&0.73&0.49&0.345&0.59\\
2(c)&\textbf{0.4}9&\textbf{0.345}&\textbf{0.59}&0.125&0.375&2.25&0.73&\textbf{0.75}&\textbf{1.375}&\textbf{1.875}\\
\end{tabular}
\end{ruledtabular}
\end{table*}

\begin{table*}[tb]
\caption{\label{tab:table1}%
Eigenvalue of Ising problems ``1/2(a)" row indicates Ising problems of Fig. 1(a) and Fig. 2(a), \textit{idem}.
}
\begin{ruledtabular}
\begin{tabular}{lccc|lccc|lccc|lccc}
\textrm{Level}&
\textrm{1/2(a)}&
\textrm{1/2(b)}&
\textrm{1/2(c)}&
\textrm{Level}&
\textrm{1/2(a)}&
\textrm{1/2(b)}&
\textrm{1/2(c)}&
\textrm{Level}&
\textrm{1/2(a)}&
\textrm{1/2(b)}&
\textrm{1/2(c)}&
\textrm{Level}&
\textrm{1/2(a)}&
\textrm{1/2(b)}&
\textrm{1/2(c)}\\
\colrule
1st & -5.5 & -3.2575 & -3.255 & 5th & -3.5 & -3.0725 & -2.265 & 9th & -0.3 & -1.2775 & -1.035 & 13th & 2.5 & 3.4025 & 3.075 \\
2nd & -4.1 & -3.2525 & -3.245 & 6th & -3.1 & -2.5925 & -2.215 & 10th & 0.5 & 0.0925 & -0.555 & 14th & 5.1 & 5.0975 & 4.595 \\
3rd & -3.9 & -3.2425 & -3.235 & 7th & -2.1 & -2.2475 & -1.945 & 11th & 1.9 & 0.7325 & -0.465 & 15th & 6.3 & 6.6175 & 5.425  \\
4th & -3.7 & -3.2225 & -3.225 & 8th & -0.5 & -2.2325 & -1.275 & 12th & 2.1 & 1.5725 & 0.715 & 16th & 8.3 & 6.8825 & 8.905 \\
\end{tabular}
\end{ruledtabular}
\end{table*}

\begin{table*}[tb]
\caption{\label{tab:table1}%
State of $4$ spins corresponding to each figure. Each state is arranged in order from the one with the lowest energy, with the ground state as the 1st.
}
\begin{ruledtabular}
\begin{tabular}{lcccccccccccccccc}
\textrm{Fig}&
\textrm{1st}&
\textrm{2nd}&
\textrm{3rd}&
\textrm{4th}&
\textrm{5th}&
\textrm{6th}&
\textrm{7th}&
\textrm{8th}&
\textrm{9th}&
\textrm{10th}&
\textrm{11th}&
\textrm{12th}&
\textrm{13th}&
\textrm{14th}&
\textrm{15th}&
\textrm{16th}\\
\colrule
1(a)&\ {-}\ {-}{\footnotesize +}{\footnotesize +}&\ {-}{\footnotesize +}\ {-}{\footnotesize +}&\ {-}{\footnotesize +}{\footnotesize +}{\footnotesize +}&{\footnotesize +}\ {-}{\footnotesize +}\ {-}&\ {-}\ {-}{\footnotesize +}\ {-}&\ {-}{\footnotesize +}{\footnotesize +}\ {-}&\ {-}\ {-}\ {-}{\footnotesize +}&{\footnotesize +}{\footnotesize +}{\footnotesize +}\ {-}&{\footnotesize +}\ {-}\ {-}{\footnotesize +}&{\footnotesize +}{\footnotesize +}\ {-}{\footnotesize +}&{\footnotesize +}\ {-}{\footnotesize +}{\footnotesize +}&{\footnotesize +}{\footnotesize +}\ {-}\ {-}&{\footnotesize +}\ {-}\ {-}\ {-}&\ {-}{\footnotesize +}\ {-}\ {-}&{\footnotesize +}{\footnotesize +}{\footnotesize +}{\footnotesize +}&\ {-}\ {-}\ {-}\ {-}\\
2(a)&\ {-}{\footnotesize +}\ {-}\ {-}&\ {-}\ {-}{\footnotesize +}\ {-}&\ {-}\ {-}\ {-}\ {-}&{\footnotesize +}\ {-}{\footnotesize +}\ {-}&\ {-}{\footnotesize +}\ {-}{\footnotesize +}&\ {-}\ {-}\ {-}{\footnotesize +}&\ {-}{\footnotesize +}{\footnotesize +}\ {-}&{\footnotesize +}{\footnotesize +}{\footnotesize +}\ {-}&{\footnotesize +}\ {-}\ {-}{\footnotesize +}&{\footnotesize +}{\footnotesize +}\ {-}{\footnotesize +}&{\footnotesize +}\ {-}{\footnotesize +}{\footnotesize +}&{\footnotesize +}{\footnotesize +}\ {-}\ {-}&{\footnotesize +}\ {-}\ {-}\ {-}&\ {-}\ {-}{\footnotesize +}{\footnotesize +}&{\footnotesize +}{\footnotesize +}{\footnotesize +}{\footnotesize +}&\ {-}{\footnotesize +}{\footnotesize +}{\footnotesize +}\\
1(b)&{\footnotesize +}\ {-}{\footnotesize +}\ {-}&\ {-}{\footnotesize +}{\footnotesize +}\ {-}&\ {-}{\footnotesize +}\ {-}{\footnotesize +}&\ {-}\ {-}\ {-}{\footnotesize +}&\ {-}\ {-}{\footnotesize +}{\footnotesize +}&\ {-}{\footnotesize +}{\footnotesize +}{\footnotesize +}&{\footnotesize +}\ {-}\ {-}{\footnotesize +}&\ {-}\ {-}{\footnotesize +}\ {-}&{\footnotesize +}{\footnotesize +}{\footnotesize +}\ {-}&{\footnotesize +}\ {-}\ {-}\ {-}&{\footnotesize +}{\footnotesize +}\ {-}{\footnotesize +}&{\footnotesize +}{\footnotesize +}\ {-}\ {-}&{\footnotesize +}\ {-}{\footnotesize +}{\footnotesize +}&\ {-}{\footnotesize +}\ {-}\ {-}&\ {-}\ {-}\ {-}\ {-}&{\footnotesize +}{\footnotesize +}{\footnotesize +}{\footnotesize +}\\
2(b)&{\footnotesize +}\ {-}{\footnotesize +}\ {-}&\ {-}\ {-}\ {-}{\footnotesize +}&\ {-}\ {-}{\footnotesize +}\ {-}&\ {-}{\footnotesize +}{\footnotesize +}\ {-}&\ {-}{\footnotesize +}\ {-}\ {-}&\ {-}\ {-}\ {-}\ {-}&{\footnotesize +}\ {-}\ {-}{\footnotesize +}&\ {-}{\footnotesize +}\ {-}{\footnotesize +}&{\footnotesize +}{\footnotesize +}{\footnotesize +}\ {-}&{\footnotesize +}\ {-}\ {-}\ {-}&{\footnotesize +}{\footnotesize +}\ {-}{\footnotesize +}&{\footnotesize +}{\footnotesize +}\ {-}\ {-}&{\footnotesize +}\ {-}{\footnotesize +}{\footnotesize +}&\ {-}\ {-}{\footnotesize +}{\footnotesize +}&\ {-}{\footnotesize +}{\footnotesize +}{\footnotesize +}&{\footnotesize +}{\footnotesize +}{\footnotesize +}{\footnotesize +}\\
1(c)&{\footnotesize +}\ {-}{\footnotesize +}\ {-}&\ {-}{\footnotesize +}\ {-}{\footnotesize +}&\ {-}{\footnotesize +}{\footnotesize +}\ {-}&\ {-}\ {-}\ {-}{\footnotesize +}&{\footnotesize +}\ {-}\ {-}{\footnotesize +}&\ {-}\ {-}{\footnotesize +}\ {-}&{\footnotesize +}\ {-}\ {-}\ {-}&{\footnotesize +}{\footnotesize +}{\footnotesize +}\ {-}&\ {-}\ {-}{\footnotesize +}{\footnotesize +}&\ {-}{\footnotesize +}{\footnotesize +}{\footnotesize +}&{\footnotesize +}{\footnotesize +}\ {-}\ {-}&{\footnotesize +}{\footnotesize +}\ {-}{\footnotesize +}&\ {-}{\footnotesize +}\ {-}\ {-}&\ {-}\ {-}\ {-}\ {-}&{\footnotesize +}\ {-}{\footnotesize +}{\footnotesize +}&{\footnotesize +}{\footnotesize +}{\footnotesize +}{\footnotesize +}\\
2(c)&{\footnotesize +}\ {-}{\footnotesize +}\ {-}&\ {-}\ {-}{\footnotesize +}\ {-}&\ {-}\ {-}\ {-}{\footnotesize +}&\ {-}{\footnotesize +}{\footnotesize +}\ {-}&{\footnotesize +}\ {-}\ {-}{\footnotesize +}&\ {-}{\footnotesize +}\ {-}{\footnotesize +}&{\footnotesize +}\ {-}\ {-}\ {-}&{\footnotesize +}{\footnotesize +}{\footnotesize +}\ {-}&\ {-}{\footnotesize +}\ {-}\ {-}&\ {-}\ {-}\ {-}\ {-}&{\footnotesize +}{\footnotesize +}\ {-}\ {-}&{\footnotesize +}{\footnotesize +}\ {-}{\footnotesize +}&\ {-}\ {-}{\footnotesize +}{\footnotesize +}&\ {-}{\footnotesize +}{\footnotesize +}{\footnotesize +}&{\footnotesize +}\ {-}{\footnotesize +}{\footnotesize +}&{\footnotesize +}{\footnotesize +}{\footnotesize +}{\footnotesize +}\\
\end{tabular}
\end{ruledtabular}
\end{table*}

\begin{table*}[tb]
\caption{\label{tab:table1}%
ELTIP $\mathcal{T}_i$  example of 3 spin Ising problem. Bold characters indicate where $\mathcal{T}_i$ affected Ising coefficients and eigenvalues.
}
\begin{ruledtabular}
\begin{tabular}{lcccccc|cccccccc}
\textrm{$\mathcal{T}_i$}&
\textrm{$J_{01}$}&
\textrm{$J_{02}$}&
\textrm{$J_{12}$}&
\textrm{$h_{0}$}&
\textrm{$h_{1}$}&
\textrm{$h_{2}$}&
\textrm{$---$}&
\textrm{$--+$}&
\textrm{$-+-$}&
\textrm{$-++$}&
\textrm{$+--$}&
\textrm{$+-+$}&
\textrm{$++-$}&
\textrm{$+++$}\\
\colrule
id&3&5&10&-7&-16&13&28&-2&-30&-48&24&14&6&8\\
$\mathcal{T}_0$&\textbf{-16}&\textbf{13}&10&-7&\textbf{3}&\textbf{5}&\textbf{6}&-2&\textbf{24}&-48&\textbf{-30}&14&\textbf{28}&6\\
$\mathcal{T}_1$&\textbf{-7}&5&\textbf{13}&\textbf{3}&-16&\textbf{10}&\textbf{14}&\textbf{24}&-30&-48&\textbf{-2}&\textbf{28}&6&8\\
$\mathcal{T}_2$&3&\textbf{-7}&\textbf{-16}&\textbf{5}&\textbf{10}&13&\textbf{-48}&\textbf{-30}&\textbf{-2}&\textbf{28}&24&14&6&8\\
\end{tabular}
\end{ruledtabular}
\end{table*}
Table 1 shows three types of $4$-bit Ising coefficients of each problem's $H$ and $\mathcal{T}_{0}[H]$ shown in each of the figures. Table 2 shows eigenvalues corresponding to cases shown in each figure. The relationship between $16$ eigenvalue order and the accompanied state are shown in Table 3. 

To clarify the effect of the transformation, the four sets of coefficients and corresponding eigenstates of the 3-bit problem are shown in Table 4, where the first row is the original coefficients of the Hamiltonian $H$ and its eigenstates, and the others are its three variants given by the transformation $\mathcal{T}_i (i=0,1,2)$ acting on $H$. As shown in Table 4, the correspondence between eigenvalues and eigenstates are permutated by the effect of the transformation $\mathcal{T}$.

Since the transformation $\mathcal{T}_i$ is swapping of $\sigma_i^z$ and ancilla spin $\sigma_n^z$ in Eq. (5) followed by elimination of $\sigma_n^z$, consecutive transformation $\mathcal{T}_{i}\mathcal{T}_{j}\mathcal{T}_{i} $ is nothing but $\mathcal{S}_{in}\mathcal{S}_{jn}\mathcal{S}_{in}=\mathcal{S}_{ij}$. 
Thus the algebraic relations of the transformation $\mathcal{T}_i$s are, (a) $\mathcal{T}_{i}\mathcal{T}_{i}=\text{id}$ which is the identity map, (b) $\mathcal{T}_{i}\mathcal{T}_{j}\mathcal{T}_{i}=\mathcal{T}_{j}\mathcal{T}_{i}\mathcal{T}_{j}=\mathcal{S}_{ij}$. The generator is to be taken in $n$ $\mathcal{T}_i$s in the case of an $n$ spin problem. 

Applying the transformation $m$ times, we have $\prod_{i=0}^{m-1}\mathcal{T}_{k_i}$, which can be simplified by applying a spin swapping $\mathcal{S}_{k_{0}k_{1}}$ from the left:
\begin{eqnarray}
\mathcal{S}_{k_{0}k_{1}}\prod_{i=0}^{m-1}\mathcal{T}_{k_{i}} &=&(\mathcal{T}_{k_{0}}\mathcal{T}_{k_{1}}\mathcal{T}_{k_{0}})\mathcal{T}_{k_{0}}\mathcal{T}_{k_{1}}\mathcal{T}_{k_{2}}\cdots\mathcal{T}_{k_{m-1}}\nonumber \\
&&=\mathcal{T}_{k_{0}}\mathcal{T}_{k_{2}}\cdots\mathcal{T}_{k_{m-1}}
\label{eq:five}.
\end{eqnarray}
in which the algebraic relation (a) and (b) are used.
By repeating this operation, most of $\mathcal{T}$s are eliminated, resulting in a single $\mathcal{T}$ or id with multiple swaps.
In such a way, generated by $\mathcal{T}_i$, we have eigenvalue invariant transformation group for Ising problem, whose order is $(n+1)!$ in which the number of the different energy landscape is at most $n+1$.

\begin{figure}[!b]
  \centering
  \includegraphics[width=1\columnwidth]{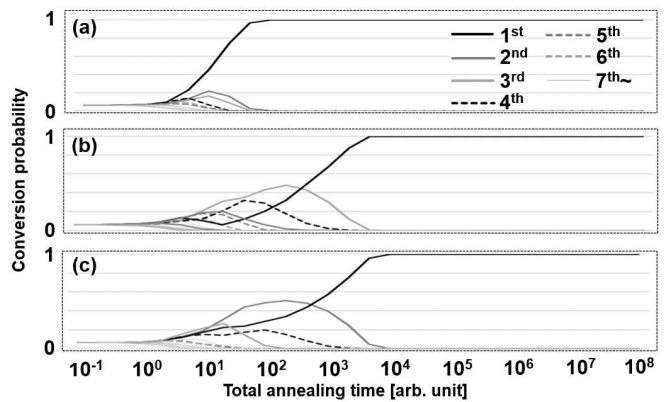}
  \caption{Simulation result of each state conversion versus QA time. Ising coefficients were swapped by $\mathcal{T}_0$, from Fig. 1(a), (b) and (c). Only small differences were seen in Fig. 2(a) comparing Fig. 1(a), however, large differences were observed between the (b) and (c) comparing the Fig. 1 and 2. }
  \label{fig:image2}
\end{figure}

Figure 2(a) and (b) show the QA simulation results of the Ising coefficients of Fig. 1 (a)and (b) with $\mathcal{T}_0$ applied. The change from application of $\mathcal{T}_0$ to the easy problem of Fig. 1(a) is small, as shown in Fig. 2(a). In Fig. 2(b), which is a difficult problem, the annealing time is about $1/2000$ times shorter than the original one. 

Looking at the eigenvalues and the state corresponding to Fig. 2(b) and Table 3, the ground state $[+-+-]$ has not changed, but the energy difference from $0.015$ to $1$ or more in the state with the longest Hamming distance $[-+-+]$ is also affected by $\mathcal{T}_0$ action. 

Figure 2(c) shows the $\mathcal{T}_0$ effect compared to Fig. 1(c). Use of ELTIP shortens the Hamming distance between the ground state and the first excited state from Hamming distance of 4 to the nearest distance of 1. The total annealing time required for convergence is almost the same in Fig. 2(b) as in 2(c), and we observe a time reduction of $10^5$ or more with $\mathcal{T}_0$. The convergence time of Fig. 1(c) is extremely long when the low energy state is at a long Hamming distance from the ground state, and there is an exponential decrease in the transition probability due to an increase of the barrier width, similar to the electron tunneling of a vacuum barrier between the electrodes.

We increased the scale of the Ising problem to 16 spins and tried real QA using a D-Wave device. The Ising coefficients were the antiferromagnetic type, whose ${J_{ij}}$s are all positive and longitudinal magnetic field term ${h_i}$ exists. We tried $20$ $\mu$sec annealing 10,000 times using the D-Wave. Success frequency of ground state convergence increased from 0 to 149 times by ELTIP. However, probably because of automatic embedding used on the D-Wave, and which may change depending on the day, it is necessary to carefully arrange the conditions for evaluating the ELTIP effect on a real QA machine.

We found that ELTIP shortens total annealing time by reducing the Hamming distance of states, whose eigenvalue is close to the ground state. If the number of spins increases for an Ising problem, the number of states close to the ground state increases exponentially. If there are many states with low eigenvalues, it is highly likely that the swapped states by ELTIP also have low eigenvalues, and the improvement of ELTIP may be reduced. However, when the real problem is embedded in the Ising problem, and if the number of the states close to the ground state in the original problem is of polynomial order, the success probability of transformed Ising problem by ELTIP is improved.

\nocite{*}

\bibliography{references}

\end{document}